\documentclass{article}
\newcommand{\ignore}[1]{} 
\usepackage{spconf,amsmath,graphicx,multirow}
\title{Sound Source Localization in a Multipath Environment Using Convolutional Neural Networks}
\twoauthors
  {\begin{tabular}{c}Eric L. Ferguson\sthanks{Work supported by Defence Science and Technology Group Australia.}, Stefan B. Williams\end{tabular}}
	{Australian Centre for Field Robotics\\
		 The University of Sydney, Australia}
  {Craig T. Jin}
	{Computing and Audio Research Laboratory\\
		 The University of Sydney, Australia}

\begin{document}
\ninept
\maketitle
\begin{abstract}
The propagation of sound in a shallow water environment is characterized by boundary reflections from the sea surface and sea floor. 
These reflections result in multiple (indirect) sound propagation paths, which can degrade the performance of\ignore{ traditional} passive sound source localization methods\ignore{ such as passive ranging by wavefront curvature}. 
This paper proposes the use of convolutional neural networks (CNNs) for the localization of sources of broadband acoustic radiated noise (such as motor vessels) in shallow water multipath environments. 
It is shown that CNNs operating on cepstrogram and generalized cross-correlogram inputs are able to more reliably estimate the instantaneous range and bearing of transiting motor vessels when the source localization performance of conventional passive ranging methods is degraded. 
The ensuing improvement in source localization performance is demonstrated using real data collected during an at-sea experiment. 

\end{abstract}
\section{Introduction}
\label{sec:intro}

Sound source localization plays an important role in array signal processing with wide applications in communication, sonar and robotics systems~\cite{Benesty2008}. 
It is a focal topic in the scientific literature on acoustic array signal processing with a continuing challenge being acoustic source localization in the presence of interfering multipath arrivals~\cite{viberg1991detection, zeng2017low, Capon1969}.
In practice, conventional passive narrowband sonar array methods involve frequency-domain beamforming of the outputs of hydrophone elements in a receiving array to detect weak signals, resolve closely-spaced sources, and estimate the direction of a sound source.
Typically, 10-100 sensors form a linear array with a uniform interelement spacing of half a wavelength at the array's design frequency. However, this narrowband approach has application over a limited band of frequencies. The upper limit is set by the design frequency, above which grating lobes form due to spatial aliasing, leading to ambiguous source directions. The lower limit is set one octave below the design frequency because at lower frequencies the directivity of the array is much reduced as the beamwidths broaden.

An alternative approach to sound source localization is to measure the time difference of arrival (TDOA) of the signal at an array of spatially distributed receivers~\cite{Carter1981, Carter1993, Chan1994, Benesty2004}, allowing the instantaneous position of the source to be estimated.
The accuracy of the source position estimates is found to be sensitive to any uncertainty in the sensor positions~\cite{ferguson2011application}. 
Furthermore, reverberation has an adverse effect on time delay estimation, which negatively impacts sound source localization~\cite{chen2005performance}. 
In a model-based approach to broadband source localization in reverberant environments, a model of the so-called early reflections (multipaths) is used to subtract the reverberation component from the signals. This decreases the bias in the source localization estimates~\cite{jensen2016}.

The approach adopted here uses a minimum number of sensors (no more than three) to localize the source, not only in bearing, but also in range. Using a single sensor, the instantaneous range of a broadband signal source is estimated using the cepstrum method~\cite{ferguson2017convolutional}. This method exploits the interaction of the direct path and multipath arrivals, which is observed in the spectrogram of the sensor output as a Lloyd’s mirror interference pattern~\cite{ferguson2017convolutional}. Generalized cross-correlation (GCC) is used to measure the TDOA of a broadband signal at a pair of sensors which enables estimations of the source bearing. Furthermore, adding another sensor so that all three sensor positions are collinear enables the source range to be estimated using the two TDOA measurements from the two adjacent sensor pairs. The range estimate corresponds to the radius of curvature of the spherical wavefront as it traverses the receiver array. This latter method is commonly referred to as passive ranging by wavefront curvature~\cite{ferguson2013modified}. However, its source localization performance can become problematic in multipath environments when there is a large number of extraneous peaks in the GCC function attributed to the presence of multipaths, and when the direct path and multipath arrivals are unresolvable (resulting in TDOA estimation bias). Also, its performance degrades as the signal source direction moves away from the array's broadside direction and completely fails at endfire. Note that this is not the case with the cepstrum method with its omnidirectional ranging performance being independent of source direction.

Recently, Deep Neural Networks (DNN) based on supervised learning methods have been applied to acoustic tasks such as speech recognition~\cite{xiao2016deep, heymann2017beamnet}, terrain classification~\cite{valada2018deep}, and source localization tasks~\cite{chakrabarty2017broadband}.
A challenge for supervised learning methods for source localization is their ability to adapt to acoustic conditions that are different from the training conditions. 
The acoustic characteristics of a shallow water environment are non-stationary with high levels of clutter, background noise, and multiple propagation paths making it a difficult environment for DNN methods.

A CNN is proposed that uses generalized cross-correlation (GCC) and cepstral feature maps as inputs to estimate both the range and bearing of an acoustic source passively in a shallow water environment.
The CNN method has an inherent advantage since it considers all GCC and cepstral values that are physically significant when estimating the source position. Other approaches involving time delay estimation typically consider only a single value (a peak) in the GCC or cepstogram.  
The CNNs are trained using real, multi-channel acoustic recordings of a surface vessel underway in a shallow water environment. CNNs operating on cepstrum or GCC feature map inputs only are also considered and their performances compared.
The proposed model is shown to localize sources with greater performance than a conventional passive sonar localization method which uses TDOA measurements.
Generalization performance of the networks is tested by ranging another vessel with different radiated noise characteristics.

The original contributions of this work are:
\begin{itemize}
\item Development of a multi-task CNN for the passive localization of acoustic broadband noise sources in a shallow water environment where the range and bearing of the source are estimated jointly;
\item Range and bearing estimates are continuous, allowing for improved resolution in position estimates when compared to other passive localization networks which use a discretized classification approach~\cite{chakrabarty2017broadband,takeda2017unsupervised};
\item A novel loss function based on localization performance, where bearing estimates are constrained for additional network regularization when training; and 
\item A unified, end-to-end network for passive localization in reverberate environments with improved performance over traditional methods.

\end{itemize}

\section{Acoustic Localization CNN}
A neural network is a machine learning technique that maps the input data to a label or continuous value through a multi-layer non-linear architecture, and has been successfully applied to applications such as image and object classification~\cite{krizhevsky2012imagenet, girshick2014rich}, hyperspectral pixel-wise classification~\cite{windrim2017} and terrain classification using acoustic sensors~\cite{valada2018deep}. CNNs learn and apply sets of filters that span small regions of the input data, enabling them to learn local correlations.

\subsection{Architecture}
\begin{figure}[t]
  \centering
\includegraphics[width=0.48\textwidth, scale=0.8]{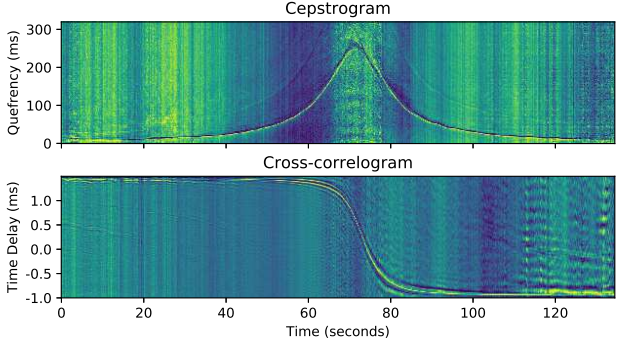}
 \caption{\textbf{a)} Cepstrogram for a surface vessel as it transits over a single recording hydrophone located 1~m above the sea floor, and 
 \textbf{b)} the corresponding cross-correlogram for a pair of hydrophones.}
\label{fig:gccandcepst}
\end{figure}

\begin{figure}[t]
  \centering
\includegraphics[width=0.48\textwidth, scale=0.4]{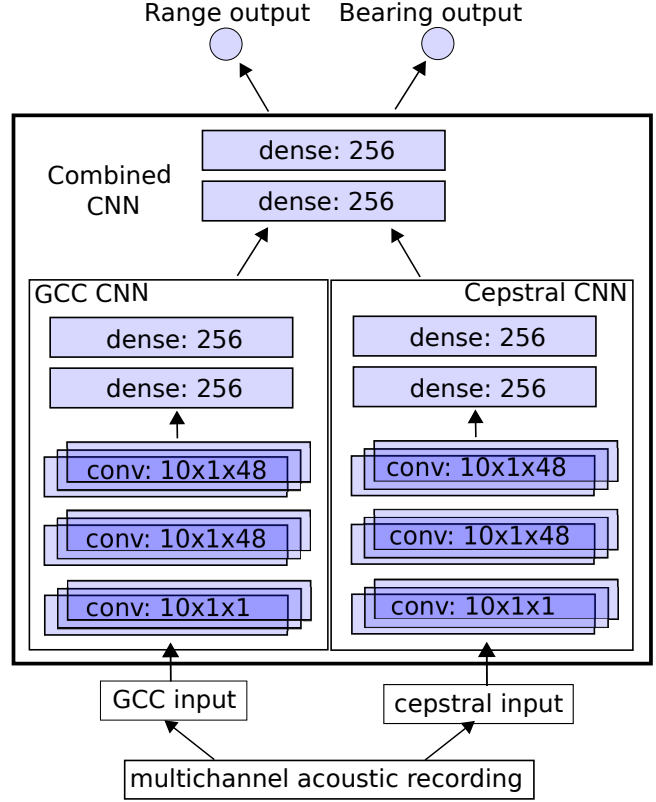}
 \caption{Network architecture for the acoustic localization CNN }
\label{fig:networkarchitecture}
\end{figure}

Since the presence of a broadband acoustic source is readily observed in a cross-correlogram and cepstrogram, Fig.~\ref{fig:gccandcepst}, it is possible to create a unified network for estimating the position of a vessel relative to a receiving hydrophone array.
The network is divided into sections, Fig \ref{fig:networkarchitecture}.
The GCC CNN and cepstral CNN operate in parallel and serve as feature extraction networks for the GCC and cepstral feature map inputs respectively.
Next, the outputs of the GCC CNN and cepstral CNN are concatenated and used as inputs for the dense layers, which outputs a range and bearing estimate.

For both the GCC CNN and cepstral CNN, the first convolutional layer filters the input feature maps with  $10 \times 1  \times 1$ kernels. The second convolutional layer takes the output of the first convolutional layer as input and filters it with $10 \times 1 \times 48$ kernels. The third layer also uses  $10 \times 1 \times 48$ kernels, and is followed by two fully-connected layers.
The combined CNN further contains two fully-connected layers that take the concatenated output vectors from both of the GCC and cepstral CNNs as input.
All the fully-connected layers have $256$ neurons each. A single neuron is used for regression output for the range and bearing outputs respectively. 
All layers use rectified linear units as activation functions. Since resolution is important for the accurate ranging of an acoustic source, max pooling is not used in the network's architecture.

\subsubsection{Input}
~\label{sec:input}
In order to localize a source using a hydrophone array, information about the time delay between signal propagation paths is required. Although such information is contained in the raw signals, it is beneficial to represent it in a way that can be readily learned by the network.

A cepstrum can be derived from various spectra such as the complex or differential spectrum. 
For the current approach, the power cepstrum is used and is derived from the power spectrum of a recorded signal.
It is closely related to the Mel-frequency cepstrum used frequently in automatic speech recognition tasks~\cite{xiao2016deep, heymann2017beamnet}, but has linearly spaced frequency bands rather than bands approximating the human auditory system's response.
The cepstral representation of the signal is neither in the time nor frequency domain, but rather, it is in the quefrency domain~\cite{bogert1963quefrency}.
Cepstral analysis is based on the principle that the logarithm of the power spectrum for a signal containing echoes has an additive periodic component due to the echoes from multi-path reflections~\cite{lo2003aircraft}. Where the original time waveform contained an echo the cepstrum will contain a peak and thus the TDOA between propagation paths of an acoustic signal can be measured by examining peaks in the cepstrum~\cite{oppenheim2004frequency}.
It is useful in the presence of strong multipath reflections found in shallow water environments, where time delay estimation methods such as GCC suffer from degraded performance~\cite{gao2008time}.
The cepstrum $\hat{x}(n)$ is obtained by the inverse Fourier transform of the logarithm of the power spectrum:
\begin{equation}
 \label{eq:cepst1}
  	\hat{x}(n) = F^{-1}\big(log|S(f)|^2\big),
\end{equation}
where $S(f)$ is the Fourier transform of a discrete time signal $x(n)$.

For a given source-sensor geometry, there is a bounded range of quefrencies useful in source localization. 
As the source-sensor separation distance decreases, the TDOA values (position of peaks in the cepstrum) will tend to a maximum value, which occurs when the source is at the closest point of approach to the sensor.
TDOA values greater than this maximum are not physically realizable and are excluded.
Cepstral values near zero are dominated by source dependent quefrencies and are also excluded. 

GCC is used to measure the TDOA of a signal at a pair of hydrophones and is useful in situations of spatially uncorrelated noise~\cite{knapp1976generalized}.
For a given array geometry, there is a bounded range on useful GCC information. For a pair of recording sensors, a zero relative time delay corresponds to a broadside source, whilst a maximum relative time delay corresponds to an endfire source. TDOA values greater than the maximum bound are not useful to the passive localization problem and are excluded~\cite{ferguson2016deep, ferguson2017convolutional}.
The windowing of CNN inputs has the added benefit of reducing the number of parameters in the network.
A cepstrogram and cross-correlogram (an ensemble of cepstrum and GCC respectively, as they vary in time) is shown in Fig.~\ref{fig:gccandcepst}. 

\subsubsection{Output}
For each example, the network predicts the range and bearing of the acoustic source as a continuous value (each with a single neuron regression output).
This differs from other recent passive localization networks which use a classification based approach such that range and bearing predictions are discretized, putting a hard limit on the resolution of estimations that the networks are able to provide~\cite{chakrabarty2017broadband, takeda2017unsupervised}.

\subsection{Multi-task Joint Training}
The objective of the network is to predict the range and bearing of an acoustic source relative to a receiving array from reverberant and noisy multi-channel input signals.
Since the localization of an acoustic source involves both a range and bearing estimate, the Euclidean distance between the network prediction and ground truth is minimized when training.
Both the range and bearing output loss components are jointly minimized using a loss function based on localization performance.
This additional regularization is expected to improve localization performance when compared to minimizing range loss and bearing loss separately.

The total objective function $E$ minimized during network training is given by the weighted sum of the polar-distance loss $E_p$ and the bearing loss $E_b$, such that:
\begin{equation}
E = \alpha E_{p} + (1-\alpha) E_{b} ,
\end{equation}
where $E_{p}$ is the L$_2$ norm of the polar distance given by:
\begin{equation}
E_{p}  = y^{2}+ t^{2} - 2 y t \cos(\theta -  \phi)
\end{equation}
and $E_{b}$ is the L$_2$ norm of the bearing loss only, given by:
\begin{equation}
E_{b}  = (\theta - \phi)^{2}
\end{equation}
with the predicted range and bearing output denoted as $t$ and $\phi$ respectively, and the true range and bearing denoted as $y$ and $\theta$ respectively.
The inclusion of the $E_{b}$ term encourages bearing predictions to be constrained to the first turn, providing additional regularization and reducing parameter weight magnitudes.
The two terms are weighted by hyper-parameter $\alpha$ so each loss term has roughly equal weight. Training uses batch normalization~\cite{ioffe2015batch} and is stopped when the validation error does not decrease appreciably per epoch. 
In order to further prevent over-fitting, regularization through a dropout rate of $50$\% is used in all fully connected layers when training~\cite{srivastava2014dropout}. 

\section{Experimental Results}
\label{sec:experiment}
Passive localization on a transiting vessel was conducted using a multi-sensor algorithmic method described in \cite{schau1987passive}, and CNNs with cepstral and/or GCC inputs.
Their performances were then compared.
The generalization ability of the networks to other broadband sources is also demonstrated by localizing an additional vessel with a different radiated noise spectrum and source level.

\subsection{Dataset}
Acoustic data of a motor boat transiting in a shallow water environment over a hydrophone array were recorded at a sampling rate of $250$~kHz. The uniform linear array (ULA) consists of three recording hydrophones with an interelement spacing of $14$~m.
Recording commenced when the vessel was inbound $500$~m from the sensor array. The vessel then transited over the array and recording was terminated when the vessel was $500$~m outbound.
The boat was equipped with a DGPS tracker, which logged its position relative to the receiving hydrophone array at $0.1$~s intervals. Bearing labels were wrapped between $0$ and $\pi$ radians, consistent with bearing estimates available from ULAs which suffer from left-right bearing ambiguity.
Twenty-three transits were recorded over a two day period. One hundred thousand training examples were randomly chosen each with a range and bearing label, such that examples uniformly distributed in range only. 
A further $5000$ labeled examples were reserved for CNN training validation.
The recordings were preprocessed as outlined in Section~\ref{sec:input}.
The networks were implemented in TensorFlow and were trained with a Momentum Optimizer using a NVIDIA GeForce GTX~770 GPU. The gradient descent was calculated for batches of $32$ training examples. The networks were trained with a learning rate of $3\times 10^{-9}$, weight decay of $1\times 10^{-5}$ and momentum of $0.9$. 
Additional recordings of the vessel were used to measure the performance of the methods. These recordings are referred to as the test dataset and contain $9980$ labeled examples.

Additional acoustic data were recorded on a different day using a different boat with different radiated noise characteristics. Acoustic recordings for each transit started when the inbound vessel was $300$~m from the array, continued during its transit over the array, and ended when the outbound vessel was $300$~m away. This dataset is referred to as the generalization set and contains $11714$ labeled examples.

\subsection{Input of Network}
\label{sec:expResinput}
Cepstral and GCC feature maps were used as inputs to the CNN and they were computed as follows. 
For any input example, only a select range of cepstral and GCC values contain relevant TDOA information and are retained - see Section \ref{sec:input}.
Cepstral values more than $1.4$~ms are discarded because they represent the maximum multipath delay and occur when the source is directly over a sensor. 
Cepstral values less than $84$~$\mu$s are discarded since they are highly source dependent. Thus, each cepstrogram input is liftered and samples $31$ through $351$ are used as input to the network only.
A cepstral feature vector is calculated for each recording channel, resulting in a $320$~x~$3$ cepstal feature map.
Due to array geometry, the maximum time delay between pairs of sensors is $\pm 9.2$~ms.
A GCC feature vector is calculated for two pairs of sensors, resulting in a $4800$~x~$2$ GCC feature map. The GCC map is further sub-sampled to size $480$~x~$2$, which reduces the number of network parameters.
\begin{figure}[t]
\centering
\includegraphics[width=0.48\textwidth, clip=true, trim=20 195 20 20 ]{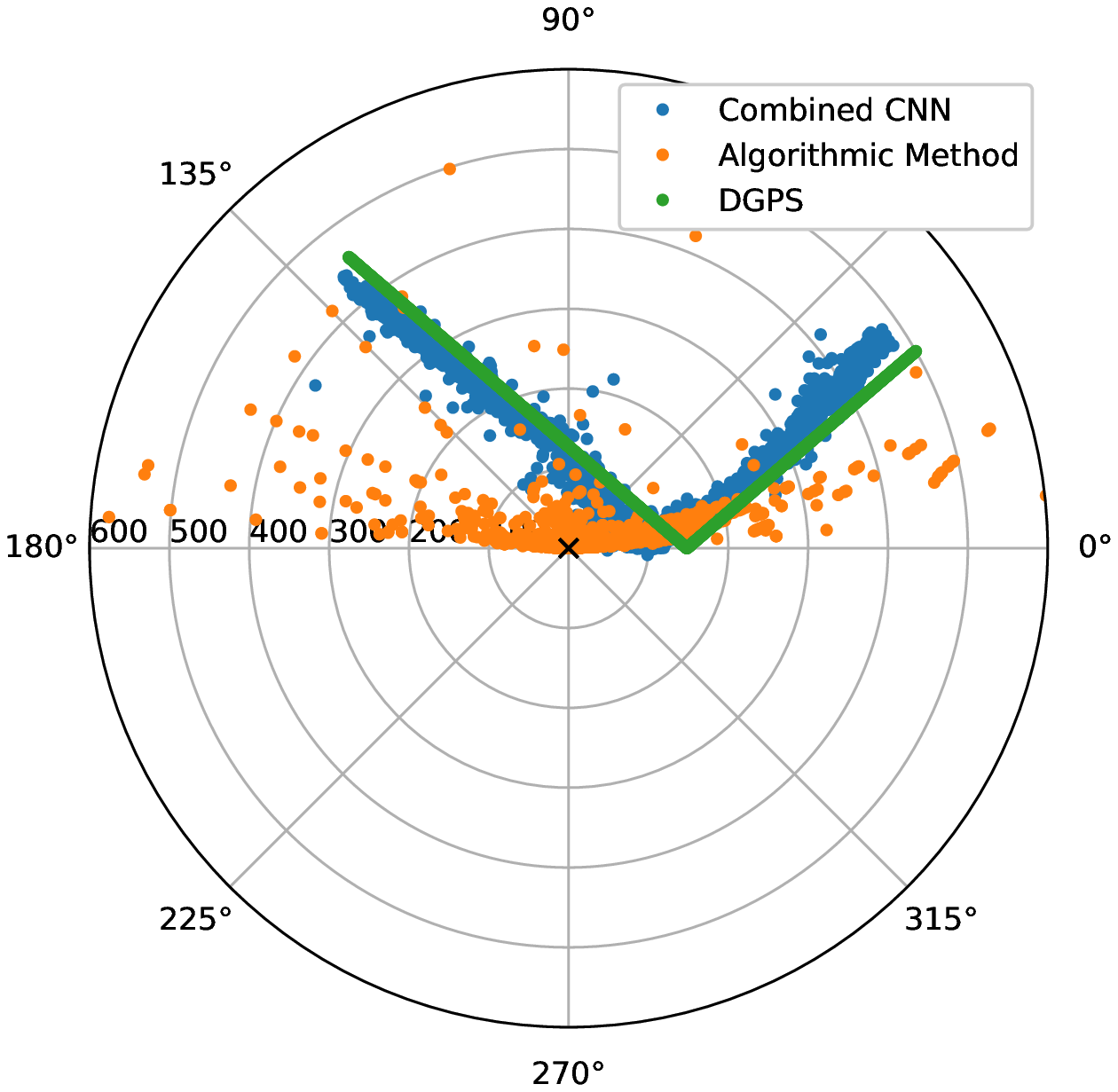}
\caption{Estimates of the range and bearing of a transiting vessel. The true position of the vessel is shown relative to the recording array, measured by the DGPS.}
\label{fig:results1}
\end{figure}

\subsection{Comparison of Localization Methods}
\begin{figure}[t]
\centering
\includegraphics[width=0.48\textwidth, clip=true, trim=20 0 30 30 ]{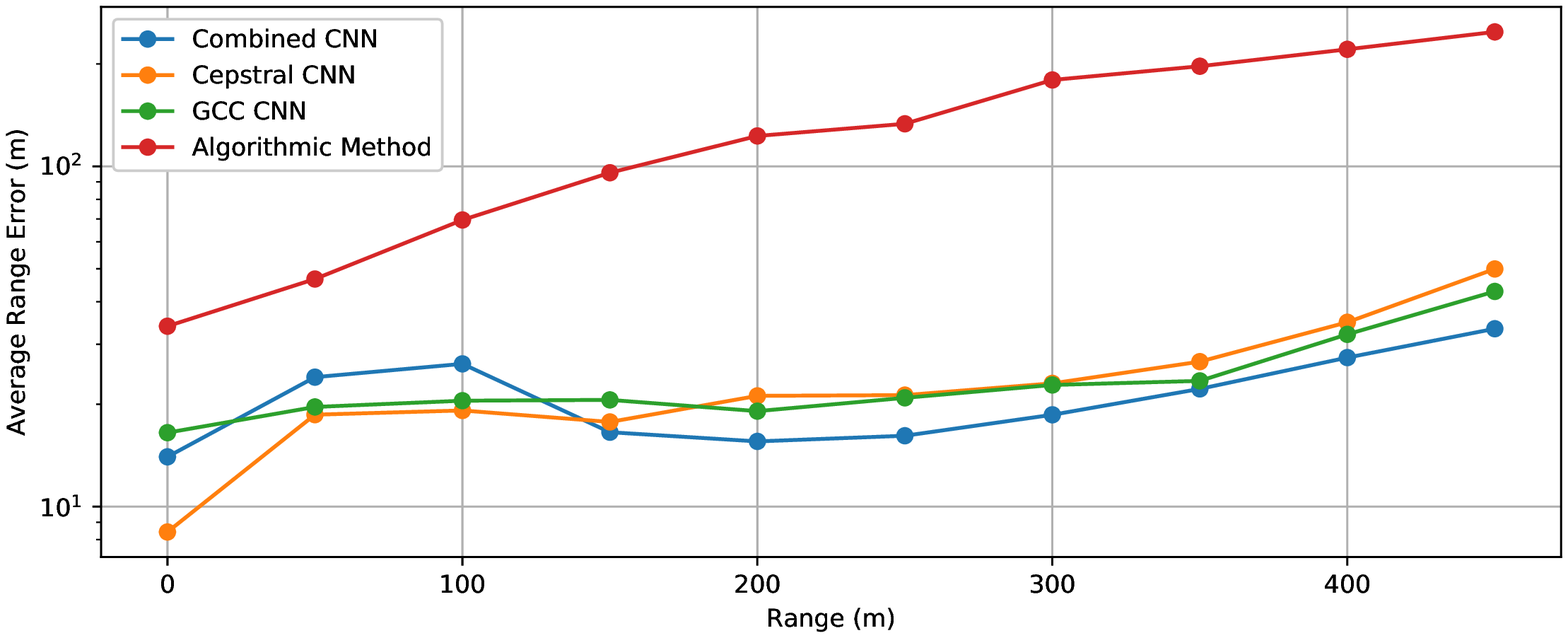}
\includegraphics[width=0.48\textwidth, clip=true, trim=20 0 30 30 ]{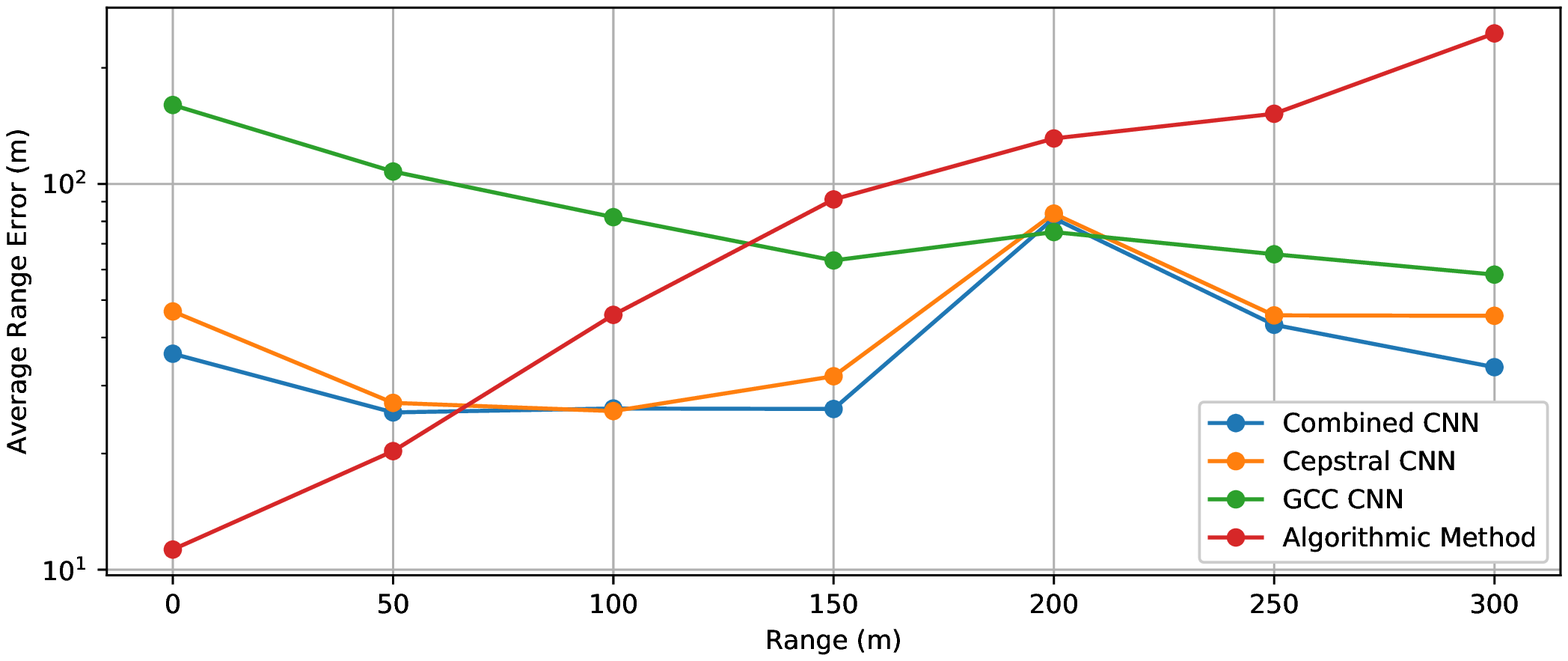}
\caption{Comparison of range estimation performance as a function of the vessels true range for the \textbf{a)} test dataset and \textbf{b)} generalization dataset.}
\label{fig:perfOverRange}
\end{figure}
\begin{figure}[t]
\centering
\includegraphics[width=0.48\textwidth, clip=true, trim=20 0 30 30 ]{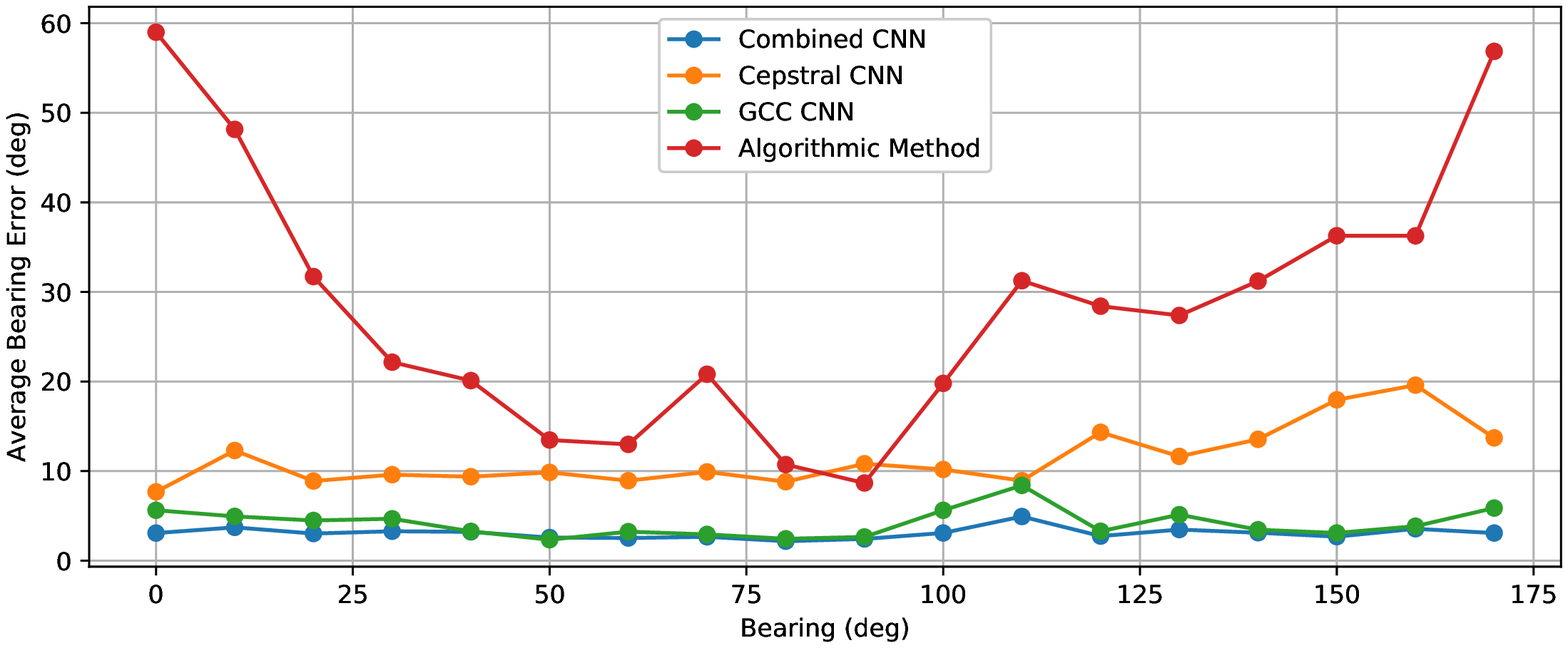}
\includegraphics[width=0.48\textwidth, clip=true, trim=20 0 30 30 ]{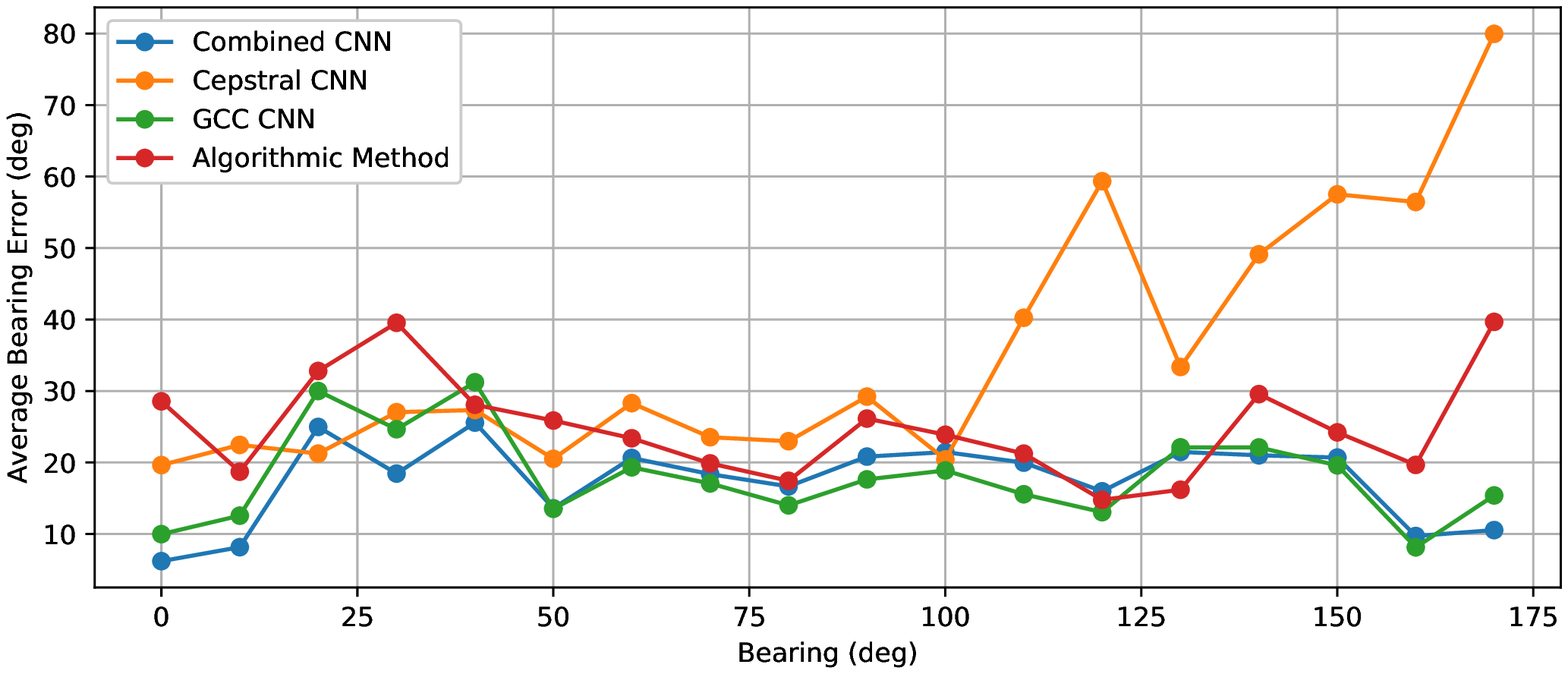}
\caption{Comparison of bearing estimation performance as a function of the vessels true bearing for the \textbf{a)} test dataset and \textbf{b)} generalization dataset.}
\label{fig:perfOverSNR}
\end{figure}

Algorithmic passive localization was conducted using the methods outlined in \cite{schau1987passive}. The TDOA values required for algorithmic localization were taken from the largest peaks in the GCC. Nonsensical results at ranges greater than 1000~m are discarded.
Other CNN architectures are also compared. The GCC CNN uses the GCC CNN section of the combined CNN only, and the Cepstral CNN uses the Cepstral CNN section of the combined CNN only, both with similar range and bearing outputs, Fig~\ref{fig:networkarchitecture}.
Fig.~\ref{fig:results1} shows localization results for a vessel during one complete transit.
Fig.~\ref{fig:perfOverRange} and Fig.~\ref{fig:perfOverSNR} show the performance of localization methods as a function of the true range and bearing of the vessel for the test dataset, and the generalization set respectively. The CNNs are able to localize a different vessel in the generalization set with some impact to performance. The performance of the algorithmic method is degraded in the shallow water environment since there are a large number of extraneous peaks in the GCC attributed to the presence of multipaths, and when the direct path and multipath arrivals become unresolvable (resulting in TDOA estimation bias).
Bearing estimation performance is improved in networks using GCC features, showing that time delay information between pairs of spatially distributed sensors is beneficial. The networks show improved robustness to interfering multipaths. 
Range estimation performance is improved in networks using cepstral features, showing that multipath information can be useful in determining the sources range.
The combined CNN is shown to provide superior performance for range and bearing estimation.

\section{Conclusions}
\label{sec:concl}
In this paper we introduce the use of a CNN for the localization of surface vessels in a shallow water environment. We show that the CNN is able to jointly estimate the range and bearing of an acoustic broadband source in the presence of interfering multipaths. 
Several CNN architectures are compared and evaluated. The networks are trained and tested using cepstral and GCC feature maps as input derived from real acoustic recordings.
Networks are trained using a novel loss function based on localization performance with additional constraining of bearing estimates.   
The inclusion of both cepstral and GCC inputs facilitates robust passive acoustic localization in reverberant environments, where other methods can suffer from degraded performance.


\vfill\pagebreak
\bibliographystyle{IEEE}
\bibliography{references.bib}

\end{document}